\documentclass{aa}  
\pdfoutput=1
\usepackage{graphicx}
\usepackage{epsfig}
\usepackage{txfonts}
\usepackage{natbib}
\begin{document}

   \title{Mapping the inner regions of the polar disk galaxy NGC~4650A\\ with MUSE\thanks{This work is based on observations taken at the ESO La Silla Paranal Observatory within the MUSE Commissioning.}}

   \author{E. Iodice\inst{1}
          \and
          L. Coccato\inst{2}
          \and
           F. Combes\inst{3,4}
          \and
          T. de Zeeuw\inst{2,5}
          \and
          M. Arnaboldi\inst{2,8}
          \and
          P. M. Weilbacher\inst{6}
          \and
          R. Bacon\inst{7}          
          \and
          H. Kuntschner\inst{2}
          \and
          M. Spavone\inst{1}
          }

   \institute{INAF-Astronomical Observatory of Capodimonte, via Moiariello 16, I-80131 Naples, Italy\\
              \email{iodice@na.astro.it}
         \and
             ESO,  Karl-Schwarzschild-Strasse 2, D-85748 Garching, Germany
         \and
             Observatoire de Paris, LERMA, PSL, CNRS, Sorbonne Univ. UPMC and College de France, F-75014, Paris, France 
         \and
             College de France, 11 Place Marcelin Berthelot, F-75005, Paris, France
          \and
            Leiden Observatory, Leiden University, Postbus 9513, 2300 RA Leiden, The Netherlands
         \and             
             Leibniz-Institut f$\ddot{u}$r Astrophysik Potsdam (AIP), An der Sternwarte 16, D-14482 Potsdam, Germany
          \and
             CRAL - Observatoire de Lyon, 9 avenue Charles Andr\'e, F-69230 Saint-Genis-Laval, France
          \and
             INAF, Osservatorio Astronomico di Torino, Strada Osservatorio 20, 10025 Pino Torinese, Italy 
            }

   \date{Received April 30, 2015; accepted August 20, 2015}

 
   \abstract{ The polar disk galaxy NGC~4650A was observed during the
     commissioning of the Multi Unit Spectroscopic Explorer (MUSE) at the ESO
     Very Large Telescope to obtain the first 2D map of
     the  velocity and velocity dispersion for both stars
     and gas.
     The new MUSE data allow  the analysis of the
     structure and kinematics towards the  central regions of
     NGC~4650A, where the two components co-exist. These regions were
     unexplored by the previous long-slit literature data available
     for this galaxy. 
     The stellar velocity field shows that there are two main
       directions of rotation, one along the host galaxy major axis
	       ($P.A.=67$~degrees) and the other along the polar disk
       ($P.A.=160$~degrees). The host galaxy has, on average, the
       typical pattern of a rotating disk, with receding velocities on
       the SW side and approaching velocities on the NE side, and a
       velocity dispersion that remains constant at all radii
       ($\sigma_{star} \sim 50-60$~km/s).  The polar disk shows a
       large amount of differential rotation from the centre  up to the
       outer regions, reaching $V\sim 100 - 120$~km/s at
       $R\sim 75$~arcsec $~\sim 16$~kpc.  Inside the host galaxy, a
       velocity gradient is measured along the photometric minor
       axis. Close to the centre, for $R\le 2$~arcsec the velocity
       profile of the gas suggests a decoupled component and
       the velocity dispersion increases up to $\sim 110$~km/s, while
       at larger distances it remains almost constant
       ($\sigma_{gas} \sim 30 - 40$~km/s). 
The extended view of NGC~4650A given by the MUSE data is a galaxy made of two perpendicular disks that  remain distinct and drive the kinematics right into the very centre of this object.    
In order to match this observed structure for NGC~4650A, we constructed a multicomponent mass model made by the combined projection of two disks.  By  comparing the observations with the 2D kinematics derived from the model, we found that the modelled mass distribution in these two disks can, on average, account for the complex kinematics revealed by the MUSE data,  also in the   central regions of the galaxy where the two components  coexist.    
This result is a strong constraint on the dynamics and formation history of this galaxy; it  further supports the idea that polar disk galaxies like NGC~4650A were formed through the accretion of material that has different angular momentum. }


\titlerunning{Two perpendicular disks in NGC~4650A} 
\authorrunning{Iodice E. et al.}

   \maketitle
%

\section{Introduction}\label{intro}
   
The polar disk galaxy NGC~4650A is  the prototype of the  
  class of polar-ring galaxies \citep[PRG, see][for a
  review]{Iod2014}. Since it is characterized by two components, the
  central host galaxy (HG) and the polar structure that rotate on
  almost perpendicular planes, it is also classified as a multispin
  galaxy \citep{Rubin1994},  a  class that includes all systems with a kinematically distinct
component of gas and/or stars with a variety of inclination angles
and radial extent with respect to the host galaxy.
Two main processes were proposed for the formation of the decoupled
component in multispin galaxies: accretion of material (gas and/or
stars) from outside (from other galaxies and/or environments) by the
pre-existing host galaxy, or merging galaxy  \citep[see][for
reviews]{Combes2014,Iod2014}.  In the cold dark matter scenario for
galaxy formation, such gravitational interactions play a fundamental
role in defining the morphology of ``normal'' galaxies, in particular in
the building up of spheroids. Thus, in this framework, the study of
multispin galaxies, both at low and high redshift, can shed light on
the main processes at work during galaxy interactions and on the
influence of the environment \citep[see][for a review]{Cons2014}.
Moreover, the existence of two orthogonal components of the angular
momentum makes the PRGs the ideal laboratory in which to derive the
3D shape of the dark matter spheroid that dominates the
gravitational potential \citep[see][for a review]{Arnaboldi2014}.

In the last ten years, a large amount of data (images
in the optical and near-infrared bands, emission and absorption line
long-slit spectroscopy and HI radio emission) has been collected for NGC~4650A. In
the combined analysis of the whole data set, by studying morphology,
kinematics, and stellar population, the structure and formation for
this fascinating galaxy was outlined very accurately.
In detail, NGC~4650A is the first PRG classified as a polar disk galaxy; in these galaxies stars and dust in the polar structure are traced inward within
the HG, down to the galaxy centre
\citep{Arn97,Iodice02,Gal02,Swaters2003}. Thus, the two components
co-exist in the inner regions.  The main properties of the polar
structure are {\it i)} the large amount of HI gas,
$10^{11}$~M$_{\odot}$, which is four times more extended than the
optical counterpart (i.e. up to 40~kpc from the centre); {\it ii)} the
differential rotation; and {\it iii)} the subsolar metallicity
$Z=0.2 Z_{\odot}$, which is constant over the whole extension of the
disk \citep{Spav10}.
Given its spheroidal shape, the HG in NGC~4650A has been classified as
a S0-like system.  This morphological classification was disputed by
\citet{Iod06} on the basis of the stellar kinematics along the
spheroid's main axes. In fact, even if this component is
rotationally supported, with a maximum rotation velocity along the
major axis of $V \simeq 80 - 100$~km/s, the velocity dispersion
remains almost constant ($\sigma \sim 65$~km/s) at all radii and along
both axes. These measurements put the HG far from the Faber-Jackson
relation for early-type galaxies \citep[see][]{Iod06, Iod2014}, which
shows that NGC~4650A has lower central velocity dispersion than that
measured for spheroids of comparable luminosity.

The structure, kinematics, and metallicity derived for NGC~4650A are
consistent with the formation of a polar disk galaxy through the
accretion of gas from outside, either from a cosmic filament or from a
gas-rich donor with a different angular momentum \citep{Spav10,
  Combes2014}.

From dynamical studies for the kinematics of NGC~4650A, the best
models predict a flattened E6-E7  dark halo, with its major axis
aligned along the plane of the polar disk itself \citep{Combes1996,
  Nap2014}.  The biggest uncertainties in these mass models reside 
{\it i)} in the limited spatial coverage of the long-slit spectra and
{\it ii)} in the contamination of the HG kinematics by the stars in
the polar disk, in the regions where the two components coexist. The
long-slit spectra cannot trace both the S-shape morphology of the
polar disk towards the centre of the galaxy \citep{Iod02} and the
warped arms at large radii \citep{Swaters2003}, thus the star and gas
motions for this component are not mapped in these regions.

Recently, \citet{Coccato2014} made a first attempt to disentangle the
contribution of the stars in the two components (HG and polar disk)
along the HG minor axis.  The main results indicate that along this
direction {\it i)} the significant rotation measured seems to be an
intrinsic property of the HG rather then an artefact caused by the
contamination of the stars in the polar disk and {\it ii)} the polar disk
shows a kinematic decoupling in the inner regions, which is
counter-rotating with respect to the outer regions and the HG.  These
findings pointed out again the complex structure of NGC~4650A in the
regions where the two components coexist and suggested that the
determination of the internal structure of NGC~4650A clearly
requires integral-field spectroscopy.

In this work we present new observations taken for NGC~4650A with the
Multi Unit Spectroscopic Explorer (MUSE) at the ESO Very Large
Telescope (VLT).  The wide wavelength range and the high spectral
resolution combined with the large field of view and fine angular
sampling of MUSE \citep{Bacon2010} allow us to derive the first
2D field of the rotation velocity and velocity
distribution for both stars and gas in NGC~4650A. Thus, we are able to
analyse the entire 2D structure (morphology and
kinematics) of both HG and polar disk, in particular in the regions
close to the galaxy centre where both components contribute to
the light distribution and kinematics.

In this paper, we adopt a distance to NGC~4650A of 43.4~Mpc and
$H_0=73$~km~s$^-1$~Mpc$^-1$. This implies that 1~arcsec$= 0.21$~kpc.

                
\section{Observations and data reduction}\label{data}

Observations of NGC 4650A were taken with the Multi Unit Spectroscopic
Explorer (MUSE) mounted on  Unit Telescope 4 (UT4) of the ESO VLT
at Paranal, Chile, during the instrument commissioning run on 8 and 12
February 2014. Observations were acquired in the wide field mode,
having a $1 \times 1$~arcmin field of view and a scale of
0.2~arcsec/pixel. The spectral range is $4800 - 9300 \AA$ and the
instrumental dispersion is about 55~km/s at $5200 \AA$
\citep{Bacon2010}.

\begin{figure*}
\centering
\includegraphics[width=18cm]{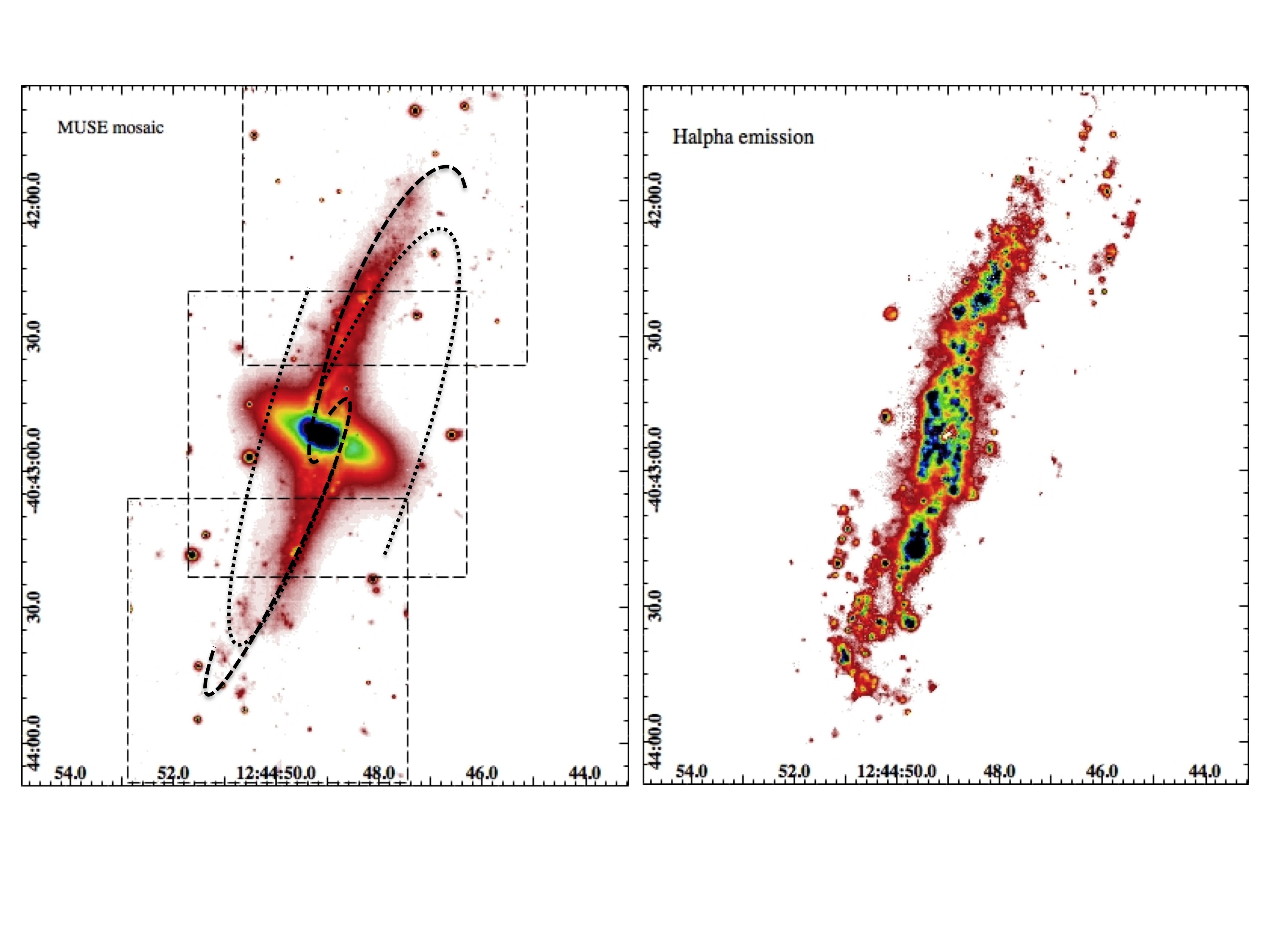} 
\caption{Left panel - MUSE mosaic of NGC~4650A in the SDSS r-band. The
  mosaic covers an area of $1.5' \times 2.5'$. The X-axis is
    the Right Ascension (hr min sec) and  the Y-axis is the
    Declination (in degrees). Dashed boxes indicate
  the single MUSE pointings.  Dashed and dotted arcs trace the
  structure of the spiral arms (see text). Right panel - H$\alpha$
  emission from MUSE data. North is up and east is on the left. }
\label{mosaic}
\end{figure*}

The data consist of 12 dithered exposures of 10 minutes each, divided
into three pointings. The central pointing, obtained by combining six
exposures, covers the host galaxy of NGC~4650A and part of the polar
structure. The northern and southern arms of the polar disk are
covered by combining three exposures for each side. A standard calibration
was adopted for this run, which includes bias, flat field, and
comparison arcs. A standard star (GD 71) was observed at the beginning
of each night to correct for instrument response across the wavelength
range. The average seeing was $FWHM \sim 0.6$~arcsec.

Data reduction was performed using the MUSE pipeline version 0.18.5
\citep{Weil2014} under the REFLEX environment \citep{Freudling+13}.
The sky background was evaluated on the regions of the field of view
where the galaxy contribution was negligible.
The different pointings were aligned using foreground stars as
reference, and then combined to create a final datacube. The alignment
consisted of offsets along the RA and DEC direction, and a $\sim 3$
degree rotation\footnote{Because the pipeline currently does not
  handle rotation when combining multiple frames, it was necessary to
  overwrite the header keywords on the raw frames accordingly.}.  We
noticed that imperfect sky subtraction left strong residuals for
wavelengths larger than $\sim$7000 \AA. We therefore concentrated our
analysis on the wavelength range 4780 \AA\ -- 6850 \AA.

The final MUSE mosaic of NGC~4650A is $1.5 \times 2.5$~arcmin (see
Fig.~\ref{mosaic}). 
 The MUSE optical r-band image of NGC~4650A (see left panel
  Fig.~\ref{mosaic}) shows clearly that the light distribution is
  dominated by two almost perpendicular disk-like components.

  The MUSE map of the $H{\alpha}$ emission is shown in the right
  panel in Fig.~\ref{mosaic}. As already known for this galaxy
  \citep{Arn97,Swaters2003}, the neutral and ionized gas component
  resides completely in the polar disk. The MUSE data reveal the whole
  2D distribution of the ionized gas and that the emission from the
  star forming regions is also quite strong  close to the centre of the galaxy.

  The MUSE data for NGC~4650A are consistent with the view given by
  previous H{\small I} data \citep{Arn97} and the optical and
  near-infrared images \citep{Iodice02,Gal02}.  They showed that the
  S-shape observed in the near-infrared images, the dust and
  bright blue knots detected in the optical data, and the H{\small I}
  iso-velocity contours are reconciled by the presence of spiral arms in
  the almost edge-on polar disk \citep{Arn97}. These features
  originate near the centre of the galaxy, the southern  arm passes in
  front of the HG and its light and dust affect the underlying HG
  component on the SW side. The structure of the spiral arms derived
  by the best fit of the H{\small I} data is plotted on the MUSE
  r-band image of NGC~4650A (see left panel in Fig.~\ref{mosaic}). The
  $H{\alpha}$ emission shows several bright knots in correspondence with
  the outer parts of the NW and SE arms.


\section{Two-dimensional kinematics of stars and gas}\label{kin}

\subsection{Measurements}

To measure the absorption line kinematics, it was necessary to
add the spectra of adjacent spatial pixels in the field of view. To
this aim, we adopted the Voronoi Binning Scheme, using the IDL
implementation by \citet{Cappellari+03}.  The ionized gas emission
lines were strong enough to be measured on individual spatial pixels,
except for the innermost $5''$, where we used the spatial bins defined
in the Voronoi Tesselation.

We first measured the stellar and ionized gas kinematics on the
spatial bins defined by the Voronoi Tesselation, using the pPXF
\citep{Cappellari2004} and Gandalf \citep{Sarzi2006} IDL
fitting routines. We then used the measured stellar kinematics to
constrain the stellar continuum in each single spatial pixel and subsequently measured
the emission line kinematics.

In both the absorption and emission line fits, we include
multiplicative polynomials of degree 8 to account for the shape of the
stellar continuum.  In the pPXF fitting, we used the library of single
stellar population spectra from \citet{Vazdekis+12}.

Each emission line was fitted independently; the mean ionized gas
kinematics was derived by averaging the results for each emission
line. The fitted emission lines are
H$\beta\lambda 4861$, [OIII]$\lambda\lambda 4959,5007$,
[NI]$\lambda\lambda 5198,5200$, [HeI]$\lambda 5876$,
[NaI]$\lambda\lambda 5890,5896$, [OII]$\lambda\lambda 6300,6364$,
[NII]$\lambda 6548$, H$\alpha\lambda 6563$, [NII]$\lambda 6583$,
[SII]$\lambda 6717$, and [SII]$\lambda 6731$.
The amplitude versus noise of each line was used as a weight in
the average;  we did not consider emission lines whose amplitude-to-noise ratio was lower than 4. The mean velocity and velocity dispersion are the luminosity
weighted mean of the lines that satisfy the above criteria.  For the
majority of the spatial bins, only the H$\alpha$, H$\beta$, [NII], and [OIII] lines
satisfy the above criteria.

\subsection{Stellar Kinematics}\label{star}

\begin{figure*}[!htbp]
\centering
\includegraphics[width=9cm]{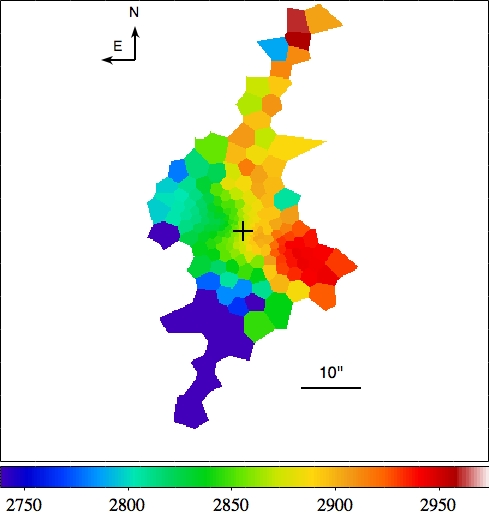} 
\includegraphics[width=9cm]{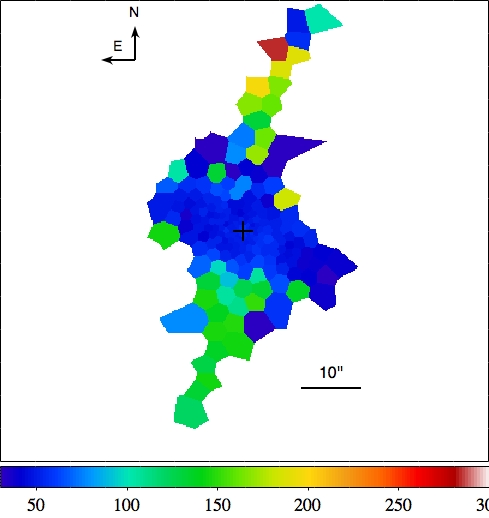} 
\caption{Map of the line-of-sight rotation velocity (left panel) and velocity dispersion (right panel) of stars in NGC~4650A. The cross marks the centre of the galaxy where the systemic velocity is $V=2875 \pm 3$~km/s and velocity dispersion is $\sigma = 60 \pm 6$~km/s.} 
\label{2Dstar}
\end{figure*}

The 2D maps of the line-of-sight velocity and
velocity dispersion for stars in NGC~4650A are shown in
Fig.~\ref{2Dstar}.  The two main directions of the star rotation are
along the HG major axis (where the position angle is $P.A.=67$~degrees) and along the polar disk
($P.A.=160$~degrees), which is almost perpendicular to the HG (see
left panel in Fig.~\ref{2Dstar}).  

Inside the HG, the velocity dispersion remains quite constant at all distances
  from the galaxy centre and P.A.s, at
  $\sigma \sim 50 - 60$~km/s, while it is larger along the polar disk
  up to $\sigma \sim 100$~km/s in the outer regions (see right panel
  in Fig.~\ref{2Dstar}).  The increase in the stellar velocity
  dispersion along the polar disk direction was already found from the
  long-slit data by \citet{Iod06}. By using this data set, after
  separating the contribution by the stars in the HG from that of the
  stars in the polar disk, \citet{Coccato2014} concluded that the
  increase of velocity dispersion observed along the polar disk is an
  intrinsic property.

 Below, we focus on the description of the 2D kinematics in the
  regions where the two components, HG and polar disk, co-exist. In
  particular, we discuss how features and perturbations in the
  velocity field of stars can be reconciled by the structure of the light
  distribution at the same radii.

\begin{figure*}[!htbp]
\centering
\includegraphics[width=9cm]{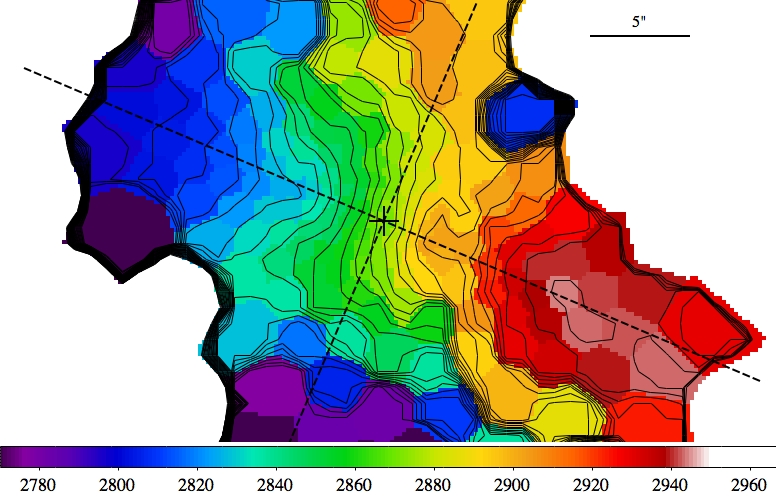} 
\includegraphics[width=9cm]{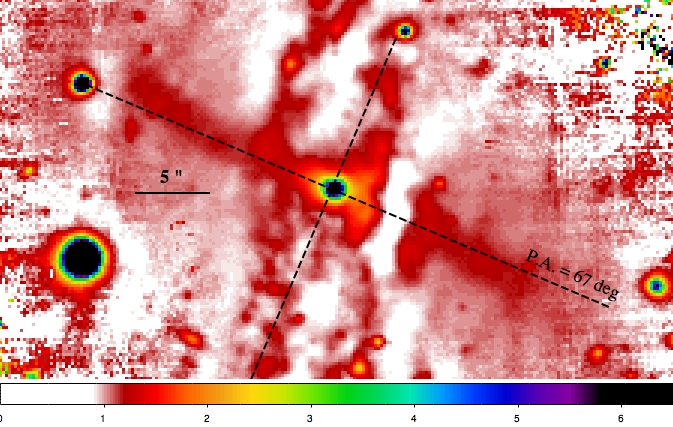} 
\includegraphics[width=9cm]{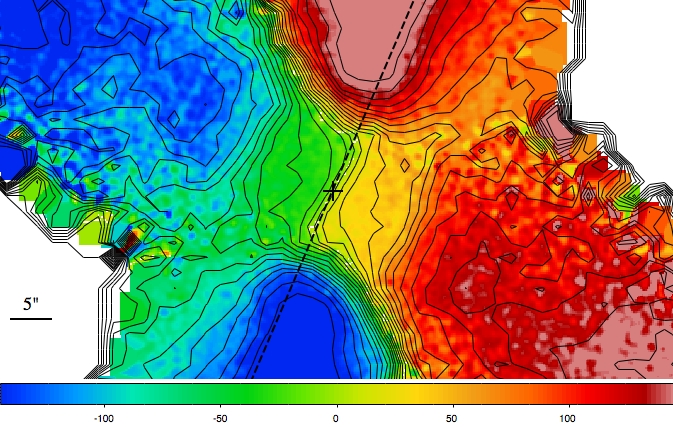} 
\includegraphics[width=9cm]{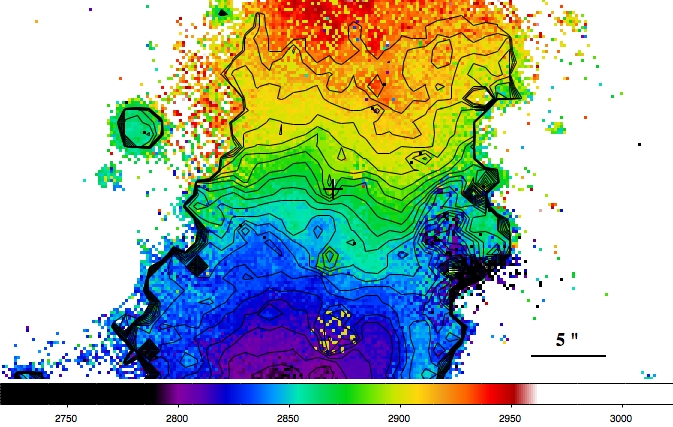} 
\caption{ {\it Top left panel}: Enlarged image of the
    2D map of the line-of-sight velocity of stars in the
    region of the HG. The iso-velocity contours are overlaid on the
    image, the minimum and maximum values are 2785~km/s and 2960~km/s,
    with a step of about 6~km/s. The cross marks the centre of the
    galaxy where the systemic velocity is $V=2875 \pm 3$~km/s. The
    dashed lines indicate the photometric major and minor axes of the
    HG. {\it Top right panel}: High-frequency residual image obtained
    by the ratio of the original reduced r-band MUSE image with a
    smoothed one, where each original pixel value is replaced with the
    median value in a rectangular window of $25\times25$~pixels. {\it
      Bottom left panel}: Enlarged image of the 2D map of
    the velocity field derived by the model. The dashed line indicates
    the direction of the HG minor axis $P.A.=157^\circ$. The
    iso-velocity contours are overlaid on the image, the minimum and
    maximum values are -145~km/s and 145~km/s, with a step of about
    13~km/s. See Sect.~\ref{model} for details.  {\it Bottom right
      panel}: Enlarged image of the 2D map of the 
      line-of-sight velocity of gas in the regions close to centre of
    the galaxy. The iso-velocity contours are overlaid on the image,
    the minimum and maximum values are 2800~km/s and 2920~km/s, with a
    step of about 6~km/s. In all images the cross marks the centre of the
    galaxy and north is up, east on the left.}
\label{zoom}
\end{figure*}

\begin{figure*}[!ht]
\centering
\includegraphics[width=9cm]{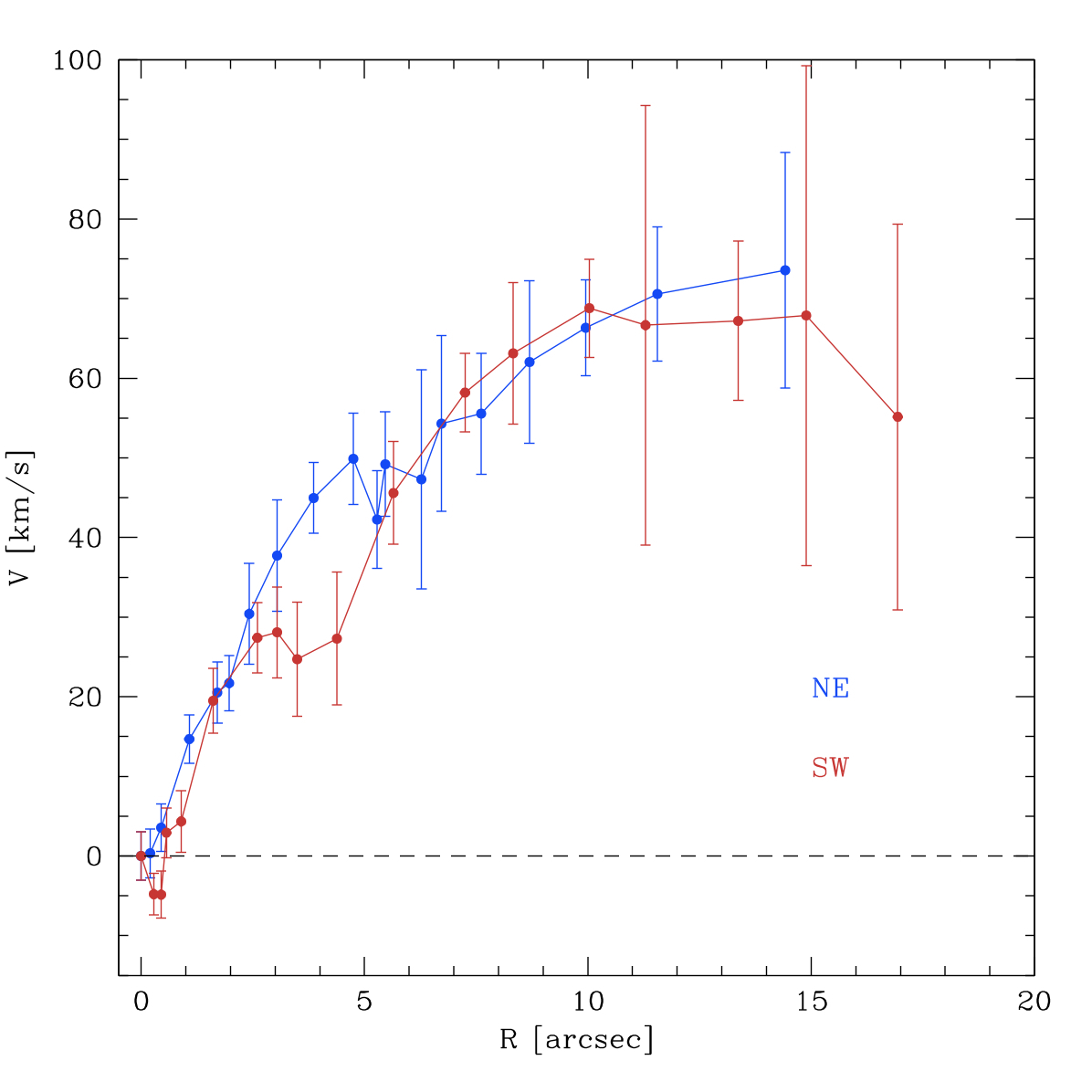} 
\includegraphics[width=9cm]{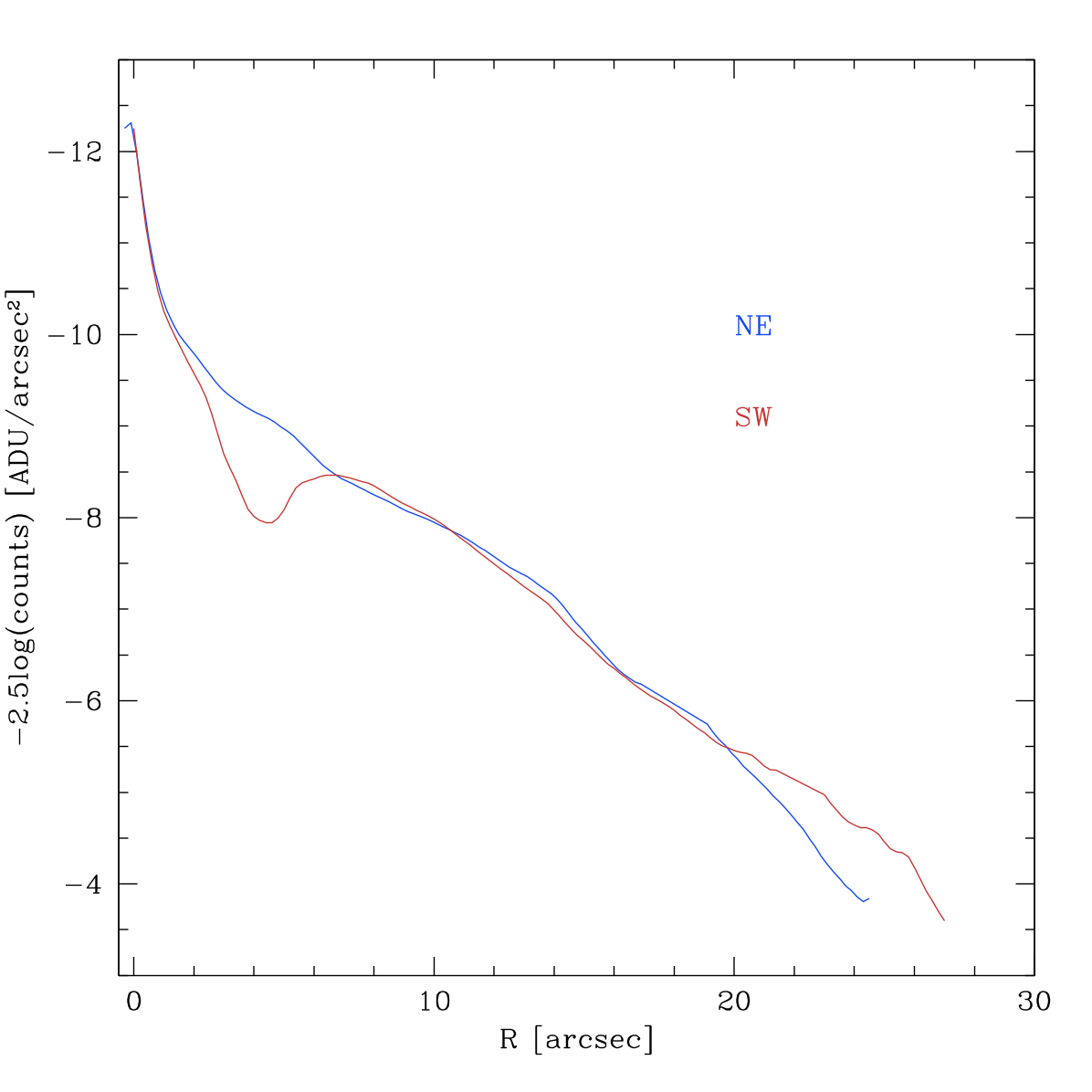} 
\caption{Folded line-of-sight velocity profiles of stars (left panel) and surface brightness profiles (right panel) along the HG photometric major axis ($P.A.=67 \pm 2$~degrees) from the MUSE data.}
\label{profiles}
\end{figure*}

  {\it HG major axis -} The overall iso-velocity contours inside the
  HG resemble the typical pattern of a rotating disk, with receding
  velocities on the SW side and approaching velocities on the NE
  side. Nevertheless, the MUSE measurements show that along the HG
  major axis the iso-velocity contours on the  side indicates a
  flatter rotation curve than the corresponding regions on the east
  side (see top left panel in Fig.~\ref{zoom}). Along this direction,
   the light distribution is also not symmetric with respect to the
  centre: the optical image shows that at $R\ge 10$~arcsec the SW
  isophotes are rounder than those at the same distance from
  centre on the other side, which appear more disky (see left panel in  Fig.~\ref{mosaic}). This asymmetry in the light is more evident from
  the high-frequency residual image\footnote{This is the ratio of the
    original reduced image with a smoothed one, where each original
    pixel value is replaced with the median value in a rectangular
    window of $25\times25$~pixels. We used the IRAF task FMEDIAN to
    smooth the original reduced image.}  of the MUSE r-band mosaic
  where the disk of the HG is more extended on the NE side
  ($R\sim20$~arcsec) than on the SW side ($R\le16$~arcsec, see the
  top left panel in Fig.~\ref{zoom}). This is a strong indication that
  the asymmetry in the kinematics reflects an asymmetry in the
  structure of the HG disk.  This effect can be quantified by looking at the folded
  light profiles and the folded line-of-sight velocity profiles
  extracted along the HG major axis, shown in Fig.~\ref{profiles}.  As
  noticed in the velocity field, the NE velocity profile  rises at
  all radii, while the SW velocities tend to have constant values for
  $R\ge 10$~arcsec. The light profiles have a different slope for
  $R\ge 16$~arcsec, where the NE profile has a steeper decline than that  the SW profile. 

  Finally, we found that the kinematic and photometric major axis are
  not coincident.  We measured the kinematic P.A. profile by using the
  software {\it Kinemetry}\footnote{Kinemetry is a generalization of
    photometry to the higher moments of the line-of-sight velocity
    distribution \citep{Kraj2006}.}, and it is shown in
  Fig.~\ref{PA}. The average value in the range of radii inside the
  regions of the HG that are not perturbed by the polar disk
  ($4 \le R \le 12$~arcsec) is $P.A._{kin}=72.2 \pm 0.6$~degrees. This
  was compared with the photometric P.A. profile derived by the fit of
  the isophote performed on the MUSE mosaic of NGC~4650A in the r-band. In the same range of radii the photometric P.A. is about 5
  degrees smaller than the kinematic P.A., being
  $P.A._{pho}=67 \pm 2$~degrees (see Fig.~\ref{PA}).

{\it HG minor axis -} Another interesting feature in the stellar
velocity field inside the HG is the non-zero velocity curve along the
P.A.=157 degrees, which corresponds to the HG minor axis and to the
polar disk major axis (see top left panel in Fig.~\ref{zoom}). This
suggests some rotation along this direction that was  already found in the velocity curves derived from the long-slit data. 
By separating the contribution of the velocity of stars in the polar disk from that in the HG, which co-exist along this direction, \citet{Coccato2014} suggested that this could be a real additional feature of the galaxy structure.

{\it Perturbations due to the polar disk -} The co-existence of
the polar disk in the inner regions of the HG (as shown by the
high-frequency residual image in the top right panel in Fig.~\ref{zoom})
generates some perturbations to the velocity distribution of stars.

Along the HG major axis, for $3 \le R \le 6$~arcsec, the iso-velocity
contours are affected by the stars in the polar disk arms which pass in
front of the HG on the SW side and behind the HG on the NE side (see
upper panel in Fig.~\ref{zoom}). The same features are also evident in the
folded light profiles and rotation curves, shown in
Fig.~\ref{profiles}.

On the NW side, at $R \sim7$~arcsec from the centre (see top left
panel in Fig.~\ref{zoom}), the velocity field shows a region where the
stellar velocity is blue-shifted with respect to those of the
surrounding area, which are red-shifted. In the same region, the
velocity dispersion is larger than the average value at the nearby
points (see right panel in Fig.~\ref{2Dstar}). We checked that these
measurements are not the result of a bad fit. They correspond to the
region of the NW arm of the polar disk that emerges from the galaxy
centre on the north side and then  goes down, passing in front of the
HG (see top right panel in Fig.~\ref{zoom}). The measured velocity and
velocity dispersion in this spatial bin, where the contribution from
the stars in the HG is negligible, have similar values to those in the
southern arm of the polar disk at about 10~arcsec from the centre.

The same explanation can be valid for another ``out-of-bound'' region
on the SE side of the HG at $R\simeq 15$~arcsec from the centre, where
velocity is lower and dispersion is higher with than in the
nearby regions (see top left panel in Fig.~\ref{zoom} and the
right panel in Fig.~\ref{2Dstar}). This area can match the kinematics
of the fainter southern arm of the polar disk which goes up and
intersects the outer regions of the HG on the east side (see left panel in 
Fig.~\ref{mosaic} and top right panel in Fig.~\ref{zoom}). In this case the velocity and velocity
dispersion are also consistent with the values measured for the SE arm of the polar disk.

\subsection{Comparison with the kinematics from long-slit data}

From the 2D kinematics of the stars in NGC~4650A we derived the
profiles of the  line-of-sight velocity and velocity dispersion,
along the major and minor axis of the HG. These are shown in
Fig.~\ref{kin_conf} and are compared with the same quantities for this
component in NGC~4650A obtained by the long-slit data taken with FORS2
at VLT \citep{Iod06} and data published by \citet{Sackett94}. The
agreement among the new and literature line-of-sight velocity
profiles is very good. The velocity dispersion measured by
\citet{Iod06} from the long-slit data is systematically larger for
measurements based on the the absorption lines around $5100 \AA$.  We
understand that this is a systematic effect caused by the
contamination of the Paschen lines in the calcium triplet (CaT)
wavelength range, which leads to a higher estimation of the velocity
dispersion \citep{Iod06}. The MUSE 2D map of the
$H\alpha$ emission also shows a strong intensity  in the regions of the
HG (see Fig.~\ref{mosaic}, right panel). This suggests the presence of
hot and young stars and hence the contamination of the Paschen lines
as the origin to the systematic effect that caused lager velocity
dispersion measurements for the CaT lines versus those measured with the 
  Mgb and Fe lines. The velocity dispersion values measured from the MUSE
spectra are derived from the Mgb and Fe lines and they are smaller.

\subsection{Gas Kinematics}\label{gas}


The 2D maps of the line-of-sight velocity and
velocity dispersion derived by the emission lines in NGC~4650A are
shown in Fig.~\ref{2Dgas}  and in the bottom right panel in Fig.~\ref{zoom}. 
 The  ionized gas kinematics is mapped out to 75~arcsec from the
  centre, i.e. up to about 16~kpc, while the previous rotation curves
  from long-slit data extend up to $10-12$~kpc \citep{Swaters2003}.
  The 2D velocity map of the ionized gas show that the polar disk has a large amount  of differential rotation from the  centre out to the outer regions, reaching $V\sim 100 - 120$~km/s at
  $R\sim 75$~arcsec $~\sim 16$~kpc. This is consistent with the  H{\small I} distribution that was well fitted by a
  model of disk in differential rotation \citep{Arn97}. 
  
  From the MUSE data of the ionized gas, the velocity of the
  approaching and receding arms on the NW and SE sides of the polar
  disk is measured for the first time. 
The gas velocity dispersion remains almost constant
($\sigma_{gas} \sim 30 - 40$~km/s) at all radii (see right panel in
Fig.~\ref{2Dgas}).

 Along the polar disk, there are some regions where the
  line-of-sight velocity and velocity dispersion have different values
   to those measured in the nearby regions. This is
  particular evident on the east side and on the outer north-west side, where
  velocities are lower (close to the systemic value) than in the adjacent
  regions. These features reflect the structure of the polar disk.  As
  discussed in Sect.~\ref{data}, the polar disk contains two
  prominent spiral arms that originate near the centre of the galaxy,
  and two fainter arms that are  visible at larger radii. The  fainter
southern  arm goes up on the east side and it overlaps the north arm at about
  30~arcsec from the centre (see Fig.~\ref{2Dgas} and also left panel
  in Fig.~\ref{mosaic}). As a consequence, in these regions the
  velocity of the gas has different values from that in the
  surrounding area. Consistently,  the velocity dispersion also has
  different values  to those in the adjacent regions.

We measured the rotation curve and velocity dispersion profile along
the polar disk major axis, at  $P.A. = 160$~degrees, and they are
shown in the left panels in Fig.~\ref{kin_gas}. The  line-of-sight
  velocity increases with distance from the galaxy centre.  It
reaches $\sim 100$~km/s at $R\sim 50$~arcsec on the NW side and
$\sim 120$~km/s at $R\sim 58$~arcsec on the SE side. The rotation
curve along the polar disk derived from the MUSE data is consistent
with that derived by \citet{Swaters2003} from the [OIII] and $H\alpha$
emission lines by using slit spectra.  The above values are
  consistent with the H{\small I} velocities in the range
  $4 \le R \le 16$~kpc\footnote{The velocities derived by the H{\small
      I} inside 4~kpc from the centre are affected by large errors
    \citep[see][]{Iodice2008}, thus values can be compared with
    velocities from the MUSE ionized gas only in the range
    $4 \le R \le 16$~kpc.}.  As  pointed out previously, the
velocity dispersion along the polar disk is almost constant
($\sigma \sim 30-40$~km/s) at all radii, except for the centre.

\begin{figure}[!htbp]
\centering
\includegraphics[width=9cm]{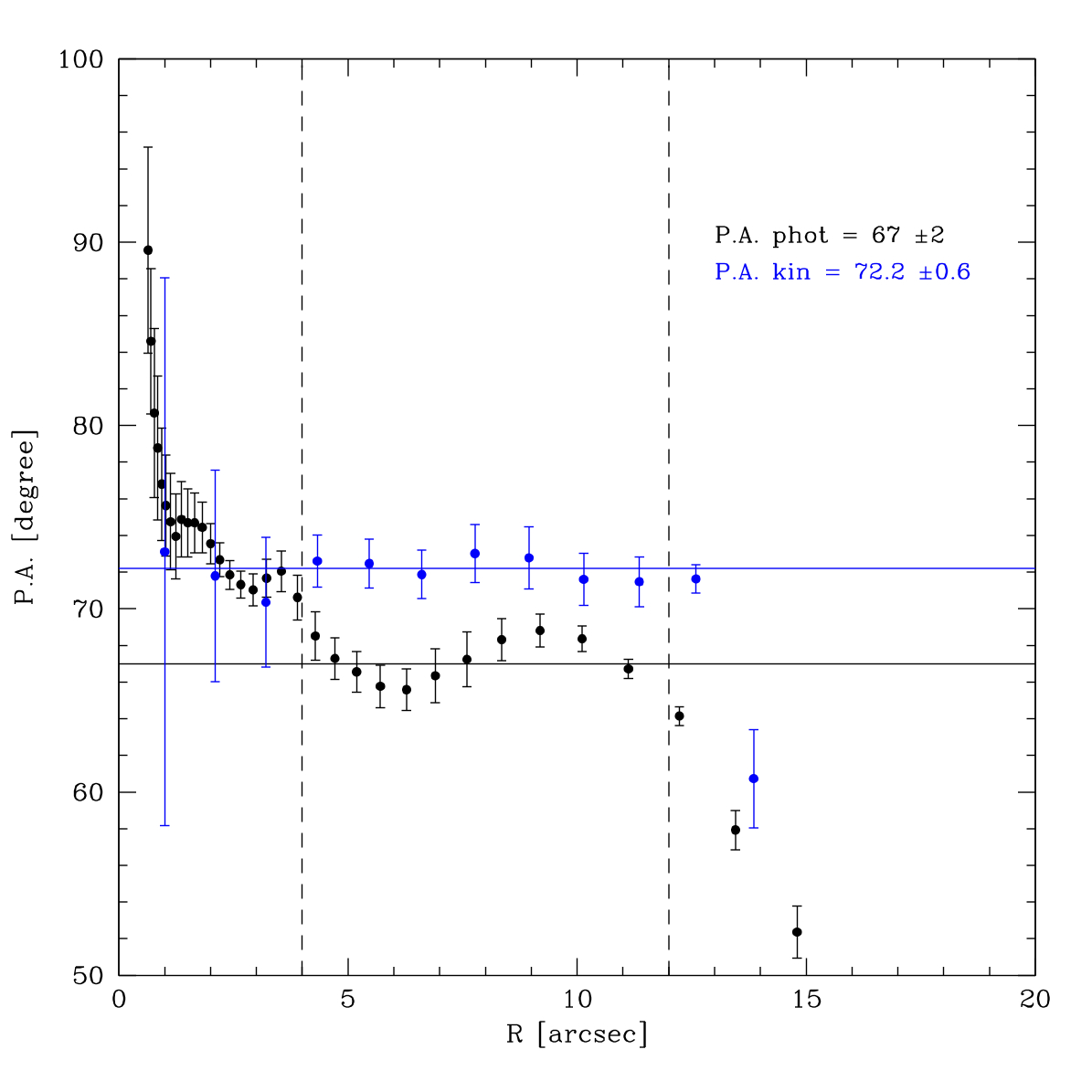} 
\caption{ Kinematic (blue points) and photometric (black points)
  P.A. profiles. The straight lines (blue and black) indicate the
  average values of the  two P.A.s in the range of radii inside
  the regions of the HG that are not perturbed by the polar disk
  ($4 \le R \le 12$~arcsec). }
\label{PA}
\end{figure}

\subsection{Gas kinematics in the galaxy centre}\label{gas_center}

The MUSE data for NGC~4650A allow us to map the 2D kinematics of the ionized 
gas all the way to the centre of this galaxy. Previous long-slit data traced
the  ionized gas kinematics only along the two arms of the polar disk where
the line emission is stronger and so did not cross the galaxy centre.

 The MUSE data show that inside 5~arcsec (see bottom right panel
  in Fig.~\ref{zoom}), the iso-velocity contours are almost parallel
  to the minor axis of the polar disk (i.e. P.A.=70 degrees),
  suggesting a cylindrical rotation inside this region. Deviations
  from such behaviour are observed closer to the centre, for
  $R\le 1$~arcsec, where the iso-velocity contours show a modest
  twist. As noted for the velocity distribution of the stars (see
  Sect.~\ref{star}), the observed distortions in the velocity field of
  the ionized gas are due to the presence of the several star forming
  regions along the polar disk, which are particularly intense when they are close to
  the galaxy centre (see right panel in Fig.~\ref{mosaic} and top right 
  panel in Fig.~\ref{zoom}).

  Inside 2~arcsec from the galaxy centre, the velocity dispersion profile 
  shows a steep increase up to a value of about 110~km/s (see upper
  right panel in Fig.~\ref{kin_gas}). In the same range of radii, a
  hint of a decoupled component  is also
  observed in the velocity curve (see bottom right panel in Fig.~\ref{kin_gas}). However, taking
  into account that the fit is based on a few pixels and that the error
  bars are large, we cannot be certain  that this feature corresponds to a
  real component.
One possible explanation of the higher velocity dispersion is the
presence in the centre of two components that lead to a broadening of
the emission line. By looking at the spectra in this region we did not
find any sign of a double peak. Thus, by summing two Gaussians
convolved with the MUSE resolution, we checked that, if two components
exist, they should have a separation in velocity of $V \le 60$~km/s;
otherwise, for  larger values a double peak should be observed in the
emission line.  Alternatively, the existence of an outflow along
  the line of sight could generate an increasing velocity dispersion
  of the gas towards the nucleus, and an almost zero velocity, as
  observed for NGC~4650A. Finally, the same effect could also be due 
  due to chaotic motions of the gas in the nuclear regions.


\section{The multicomponent mass model for NGC~4650A}\label{model}

In this section we present the numerical model made to
reproduce the observed structure and the 2D kinematics of
stars and gas showed by the MUSE data for NGC~4650A.  

 Photometry and spectroscopy (including new MUSE data) have shown
  that the light distribution and kinematics are consistent with two
  almost perpendicular disks where the two main directions of the
  star rotation are along the equatorial host disk ($P.A.=67$~degrees)
  and along the polar disk ($P.A.=160$~degrees). The two components  co-exist in the central regions of the galaxy since stars and dust
  in the polar disk can be traced up to the galaxy centre, following a
  spiral pattern (see top left panel in Fig.~\ref{zoom}).

Even if the morphology of the central HG resembles that of an S0
galaxy (see left panel in Fig.~\ref{mosaic}), the observed light
distribution \citep{Iod02} and the constant velocity dispersion of
stars at all radii (as revealed by FORS2 and the current MUSE
absorption line spectra) strongly suggested that this component can be
considered  a puff-up disk, rather than a hot spheroid.  The disk
nature of the HG is also evident  from the high-frequency
structure along the equatorial direction found in the residual image
shown in the top left panel in Fig.~\ref{zoom}.

\begin{figure*}[!htbp]
\centering
\includegraphics[width=9cm]{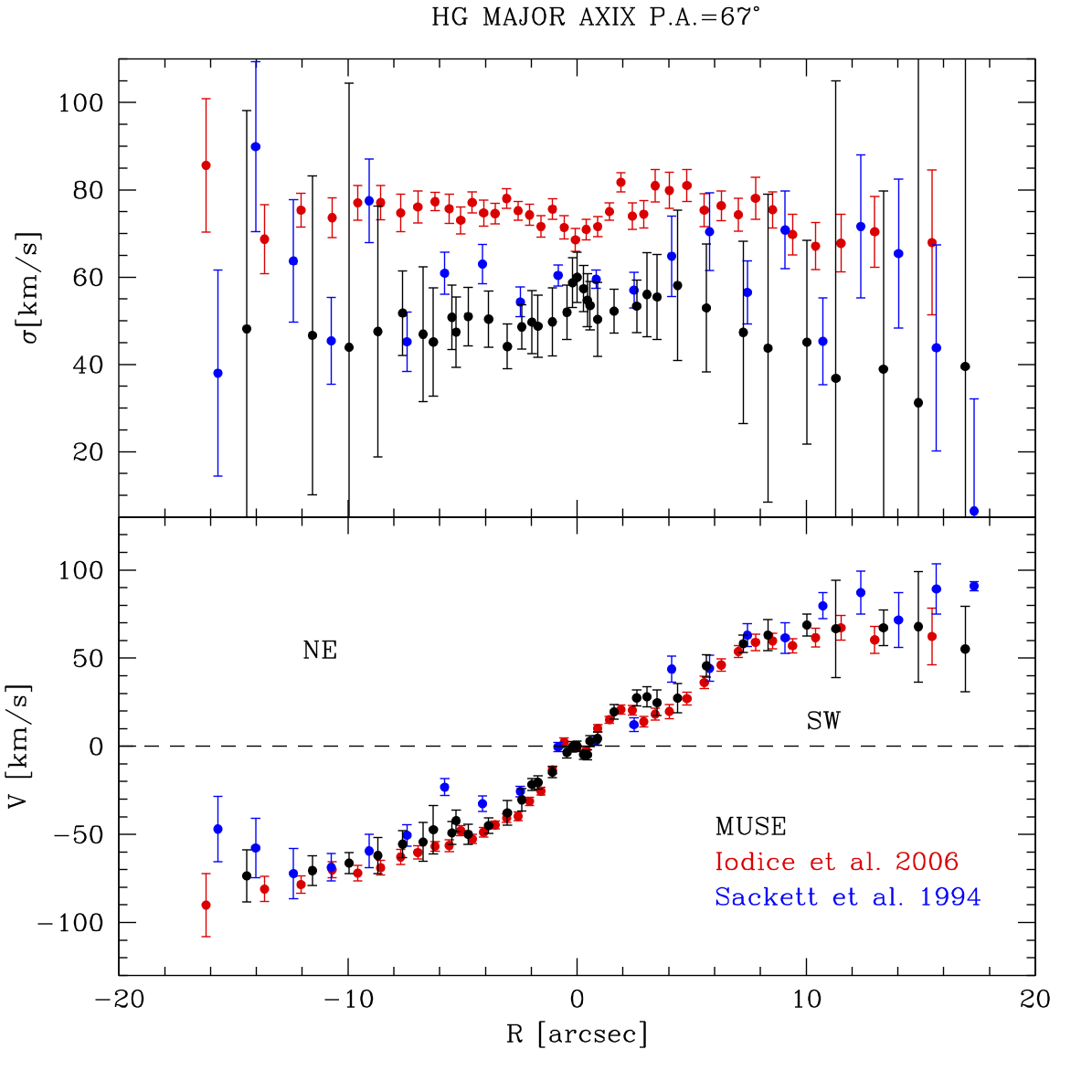} 
\includegraphics[width=9cm]{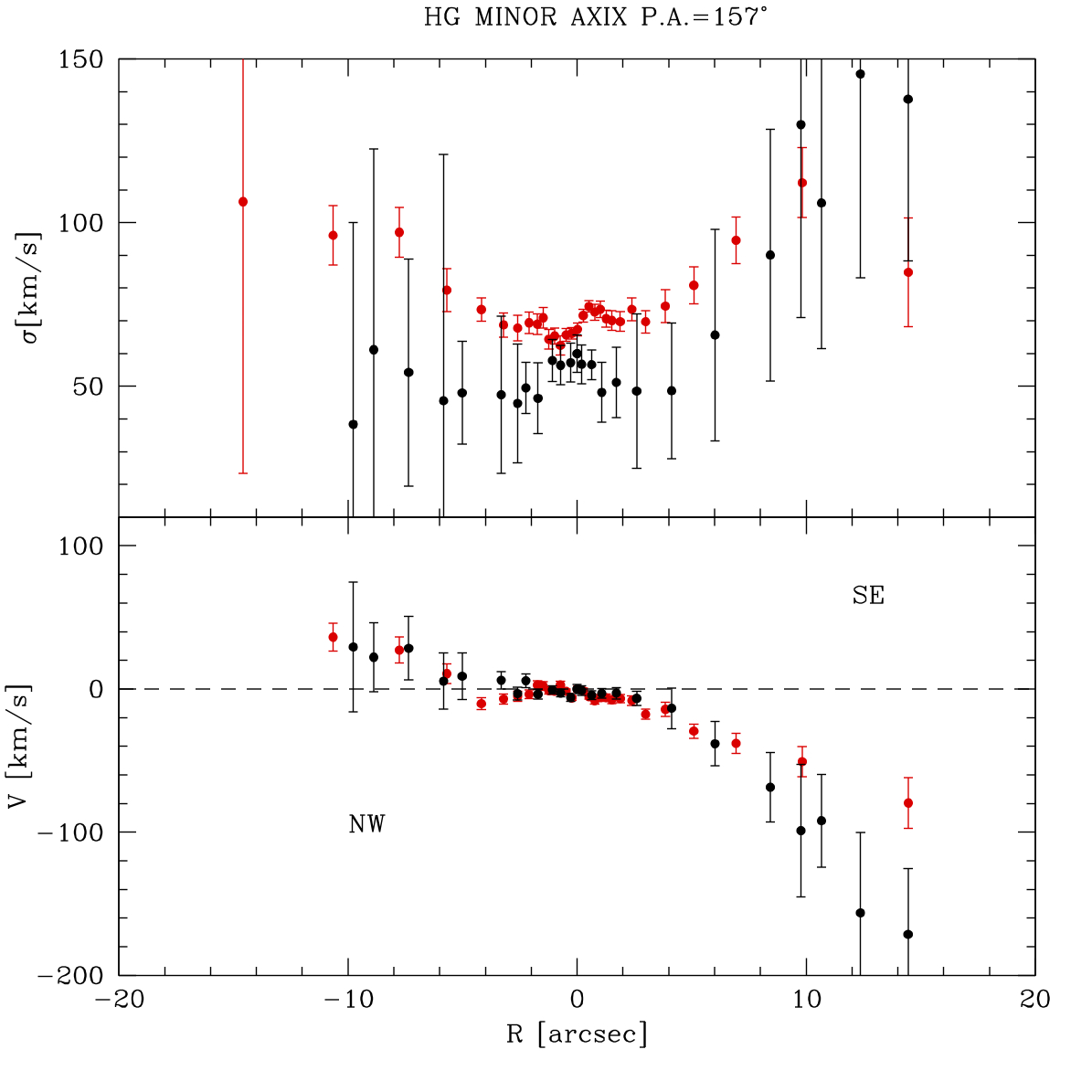} 
\caption{Stellar kinematics along the major axis (left panel) and minor
  axis (right panel) of the HG in NGC~4650A derived by MUSE data
  (black points) in the Mgb wavelength region compared with the
  profiles obtained by \citet{Sackett94}, blue points, in the same
  region, and longslit spectra by \citet{Iod06}, red points, in the
  CaT wavelength region.}
\label{kin_conf}
\end{figure*}

Based on these observational facts and taking into account that the
contribution to the total  velocity comes from  stars in
the HG and in the polar structure, we built a  minimum model made  of  two 
perpendicular disks. The low value measured for the velocity
dispersion of both components further  supports this idea.

 The model does not include any rotation along the minor axis of
  the HG, the spiral arms in the polar disk, and a decoupled component in the centre, as well as the
  misalignment between kinematic and photometric major axis of the HG,
  also revealed by the MUSE data. This approach allows us to verify
  whether the minor axis rotation in the HG and/or the decoupled
  central component are due to the projection  of the orbits in the polar disk onto the HG ones.
  If this is the case, we should also
  measure these features  in the velocity and velocity dispersion fields
  derived by the model. Alternatively, they are to be considered
  intrinsic properties of the HG in NGC~4650A.

To derive the velocity field in both equatorial and polar planes, one
technique is to build a 3D gravitational potential by combining
density models of the host galaxy and the polar disk, and then to search
for the symmetric close orbits in both the equatorial and
perpendicular planes by shooting test particles
\citep[e.g.][]{Sackett1990, Sackett94, Combes1996}. This procedure is
quite long, since particles are launched tangentially on the major
axis at each radius of the minor axis, and the launching velocity is
adjusted until the closed orbit is found.  Instead, we adopt here the
epicyclic approximation in each plane (equatorial and polar) to
compute the closed orbits in a non-axisymmetric potential
\citep[e.g.][]{Binney1987}. Indeed, in their respective planes, the
existence of the perpendicular component is equivalent to a weak bar,
with zero pattern speed.

\begin{figure*}[!htbp]
\centering
\includegraphics[width=18cm]{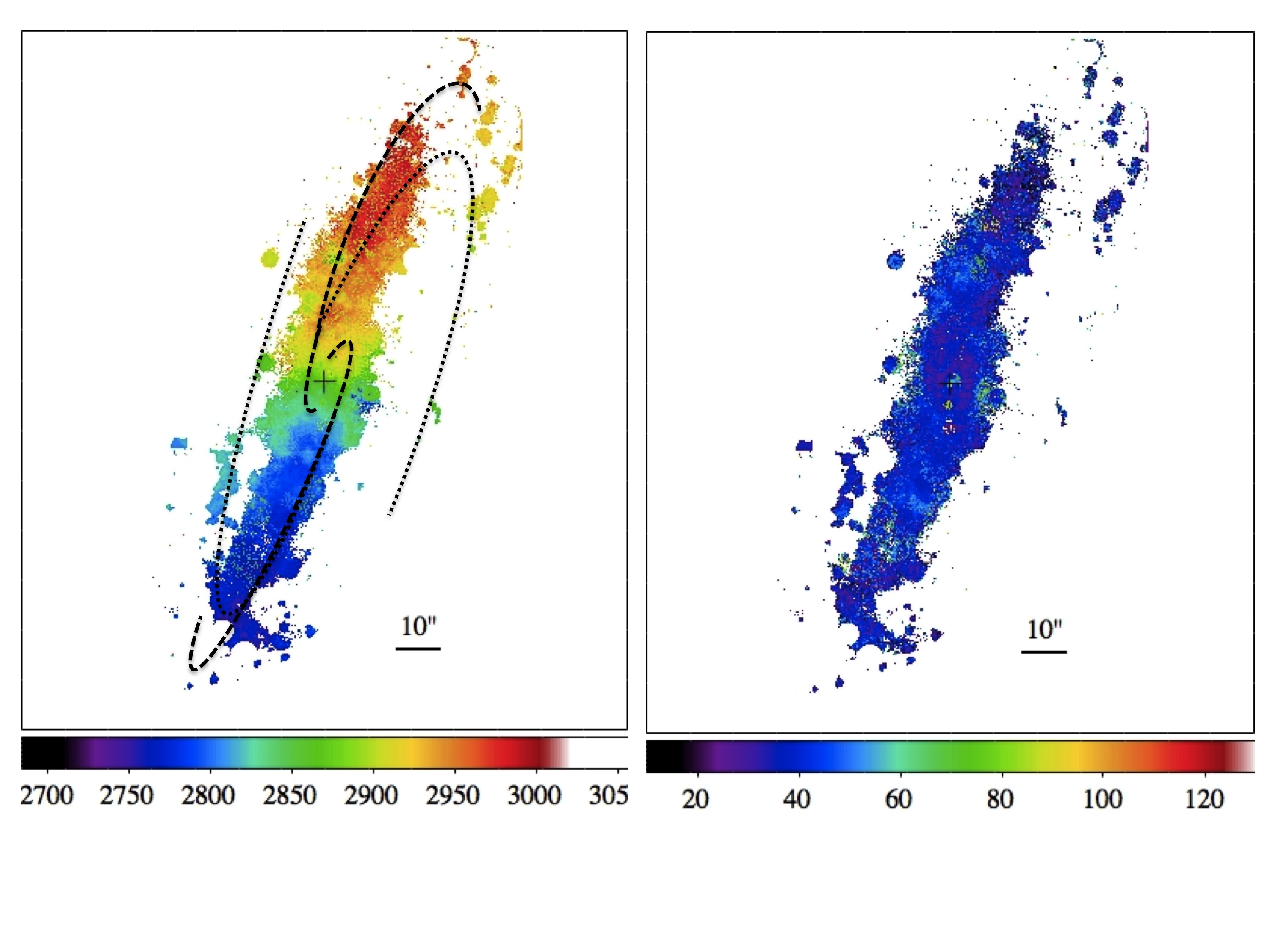} 
\caption{Maps of the  line-of-sight velocity (left panel) and
  velocity dispersion (right panel) of the gas in NGC~4650A. The cross
  marks the centre of the galaxy where rotation is 
    $V=2880 \pm 19$~km/s and velocity dispersion is
    $\sigma = 100 \pm 20$~km/s. In the left panel, the dashed and
  dotted arcs trace the structure of the spiral arms, as in Fig.~\ref{mosaic}.}
\label{2Dgas}
\end{figure*}

 In this work, we choose the multicomponent mass model previously developed by
  \citet{Combes1996}. This model was proposed in order to revisit the 3D shape of the
  dark matter halo in NGC~4650A, and they found that the luminous
  mass closely fitted  the kinematics along the major axis of the HG
  galaxy and the optical rotation curve of the polar disk, while to account  for the high polar velocities at large radii given by the H{\small I} data, a dark matter halo is required from $8-10$~kpc. The main result of this study was on the shape of the dark matter halo: the best fit to data with the least amount of dark matter was obtained with
  a flattened E6-E7 dark halo, having its major axis aligned along the  polar disk major axis.

  In this work we adopt the same configuration for NGC~4650A as in
  the model proposed by \citet{Combes1996}, i.e. two luminous components HG and a polar
  disk plus a flattened dark matter halo, with only slight
  modifications to take into account the extended and deeper photometry in the I and K bands
  published by \citet{Iod02}.

As in \citet{Combes1996}, according to the light distribution, the
central host galaxy can be described as a very small bulge plus  a
  thick disk. The bulge is very light, and its influence might be
ignored. It is represented here by a Plummer component, having a
characteristic radius  $r_B$ and a total mass $M_B$.  The host
disk is represented by a double exponential, with mass $M_d$,
characteristic scale $r_d$, and height $h_d$. All the above parameters
are given in Table \ref{tab:model}.

The stellar and gaseous polar structure are represented by different
Miyamoto-Nagai potential-density pairs \citep{MN1975}. Table
\ref{tab:model} indicates their masses $M_{PD}$, their common height
$h$, and their two internal and external radii $r_1$ and $r_2$. 
  The model of the polar disk does not account for the observed spiral
  arms and for the several star forming regions that strongly perturb
  this component.

A dark matter halo is added to account for the  high
  velocities in the outer parts. Its density shape is that of a
pseudo-isothermal ellipsoid

$$ \rho_h = \rho_0 \Big[1 + \frac{(R^2 + z^2/q^2)}{r_h^2}\Big]^{-1} $$

\noindent whose characteristic parameters are also listed in Table
\ref{tab:model}.   The values of the central density $\rho_0$ and
  the halo flattening $q$ were those derived by the dynamical model
  from \citet{Combes1996}.

The sum of all these mass components yields a 3D
potential in the whole space.  In particular, we are interested  in the
resulting closed orbits in the two planes, equatorial and polar, which
have lost their axisymmetry.  In each plane, the potential is
computed, and decomposed in Fourier components
$ \Phi(R,\phi) = \Phi_0(R) + \Phi_m(R) cos(m\phi -\phi_m) $.  We
consider only the $m=2,4,$ and $6$ components.

Following the epicyclic approximation, the closed orbits in a
non-axisymmetric potential, such as a weak bar, can be developed in
polar coordinates ($R, \phi$) as in \citet{Binney1987} and \citet{Schoenmakers1997}:

$$R(t) = R_0 + R_1(t) \quad, \quad  \phi(t) =  \phi_0(t) +  \phi_1(t)$$

\noindent We take  $\phi=0$ as the long axis of the potential.
For the unperturbed circular orbit, 

$$ \phi_0(t) = \Omega_0 t  $$

$$ R_1(\phi_0) = C \cos(m\phi_0),$$

\noindent where $m$ is the Fourier component in which
the potential has been decomposed,

$$ C = \frac{-1}{\Delta} \big(\frac{d\Phi_m}{dR} + 2 \frac{\Phi_m}{R}\big)_{R_0}, $$ 

\noindent and 

$$ \Delta = \kappa_0^2 - m^2 \Omega_o^2 $$

\noindent with $\kappa_0$ the unperturbed epicyclic frequency.

\noindent The expression of the perturbed angle is

$$ \phi_1 =  \frac{\sin(m\phi_0)}{m}  \Big[  \frac{2}{\Delta R_0} \big(\frac{d\Phi_m}{dR} + 2 \frac{\Phi_m}{R}\big) - \frac{\Phi_m}{R_0^2\Omega_0^2} \Big]. $$
 
\noindent Computing derivatives of these quantities, the velocities in the plane of the closed
orbits can be obtained as

$$ V_R = m V_0 \frac{1}{\Delta R_0}  \big(\frac{d\Phi_m}{dR} + 2 \frac{\Phi_m}{R}\big) \sin(m\phi_0), $$

$$ V_{\phi} = V_0  +  V_0 \cos(m\phi_0) \Big[ \frac{1}{\Delta R_0}  \big(\frac{d\Phi_m}{dR} + 2 \frac{\Phi_m}{R}\big)  - \frac{\Phi_m}{R_0^2\Omega_0^2} \Big]. $$

\begin{figure*}[!htbp]
\centering
\includegraphics[width=15cm]{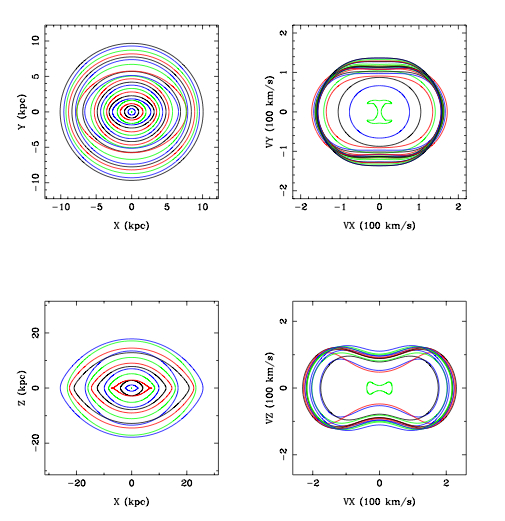} 
\caption{Closed orbit positions (left panels) and velocities (right) for the NGC~4650A
host galaxy (top) and polar structure  (bottom). }
\label{orbpol}
\end{figure*}

\noindent To sum the potential of the various components, we adopt a
grid of 480x480 in each plane, equatorial and polar.  The system of
closed orbits found in the equatorial and polar planes are displayed
in Fig.~\ref{orbpol}.

\begin{figure*}[!htbp]
\centering
\includegraphics[width=9cm]{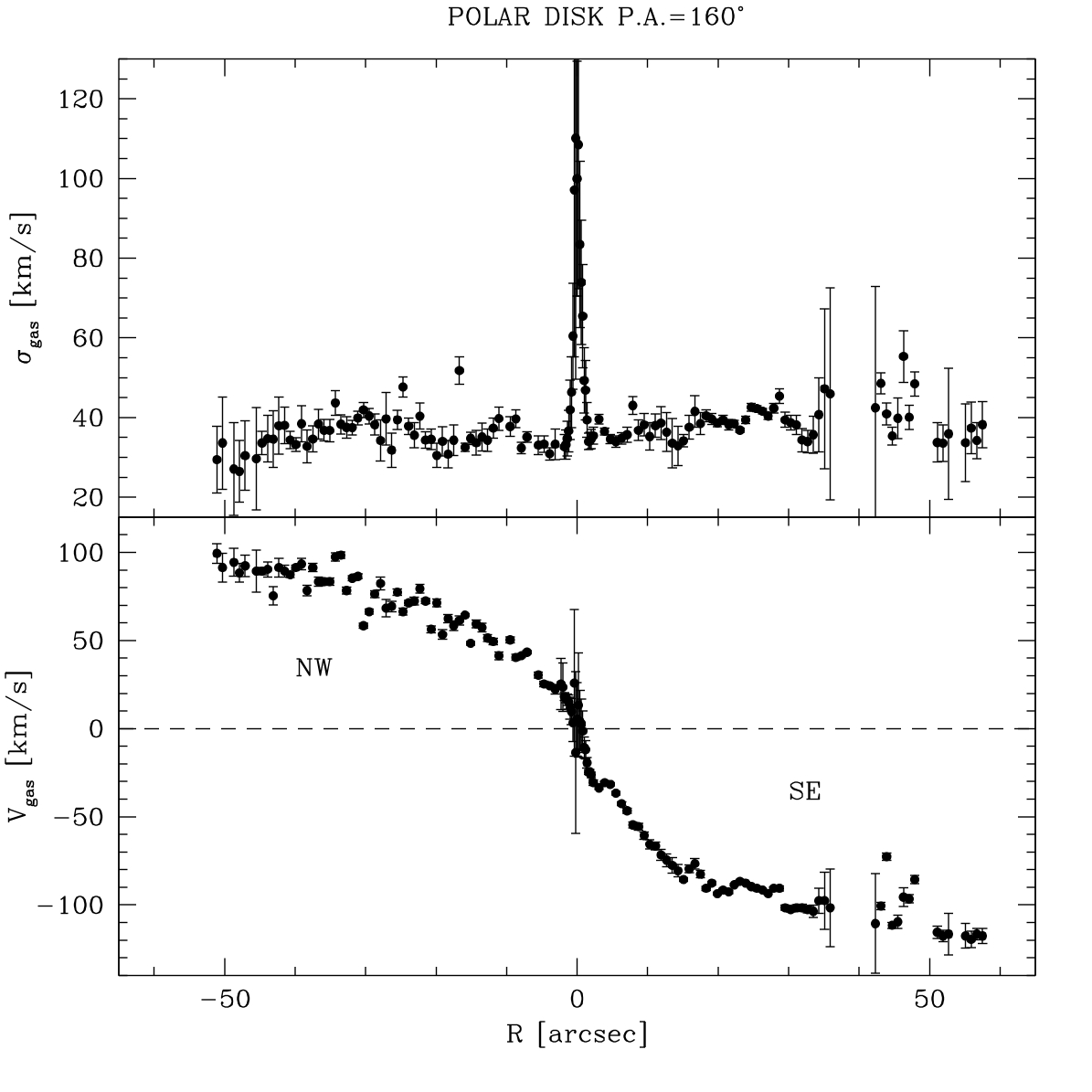} 
\includegraphics[width=9cm]{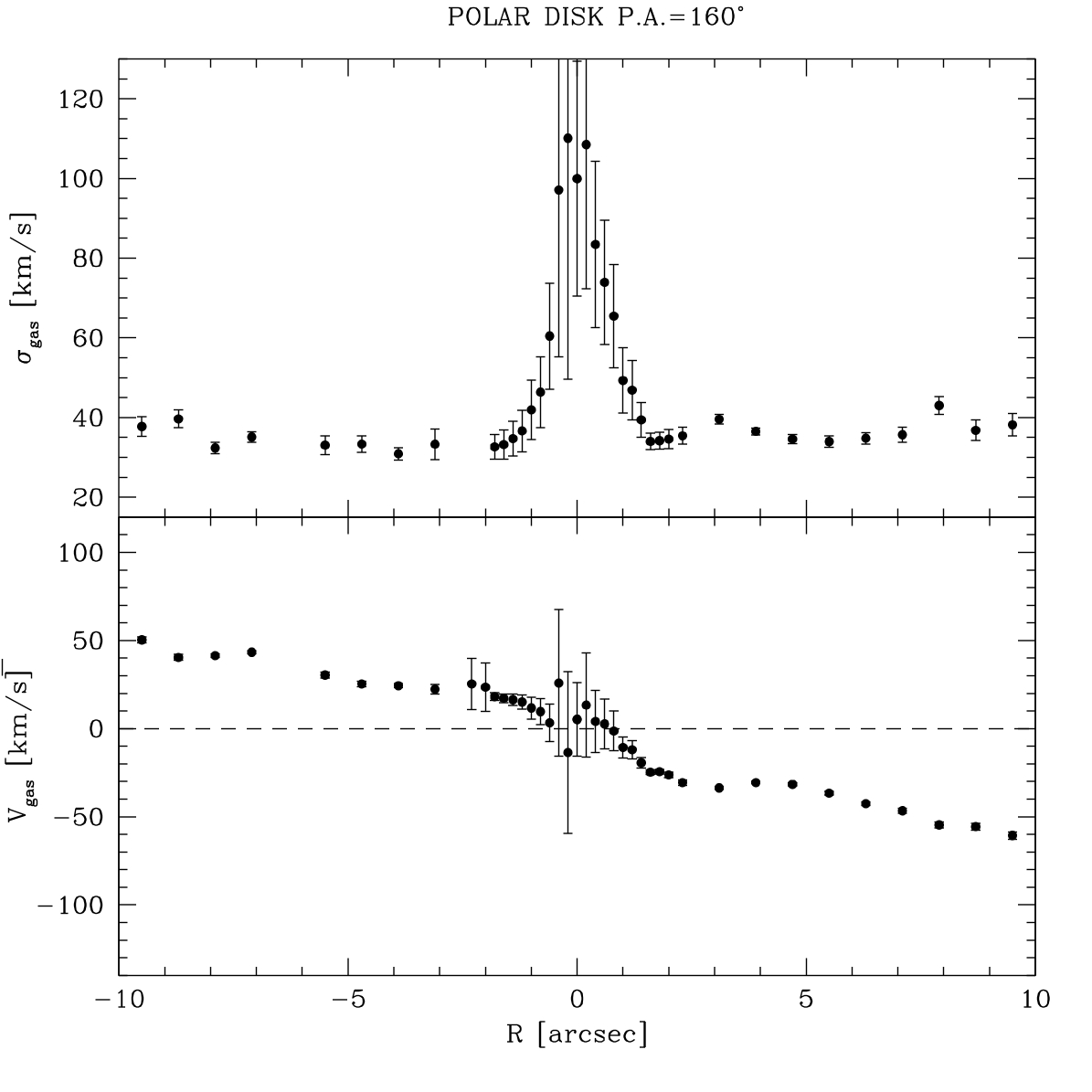} 
\caption{Left panel -  Line-of-sight velocity (bottom panel) and
  velocity dispersion (upper panel) of the gas along the polar disk
  major axis, $P.A.= 160^{\circ}$. Right panel - Zoom of the gas
  rotation (bottom) and dispersion (upper) curves towards the centre
  of the galaxy.}
\label{kin_gas}
\end{figure*}

We then undertake to project the two components, host galaxy and polar
disk, taking  their thickness into account, with the adopted shapes and
parameters indicated in Table \ref{tab:model}.  As for the velocities,
we take  velocity dispersions of the stars into account, as well as their
asymmetric drift, i.e. we adopt for an axisymmetric exponential disk
(with scalelength $r_d$) the Jeans equation,

$$ v_c^2-v_\phi^2 = \sigma_\phi^2-\sigma_r^2(1-r/r_d) -\frac{\partial\sigma_r^2}{\partial lnr},$$

\noindent where the dispersion $\sigma_r$ and $\sigma_\phi$ in the radial and
tangential directions are related in the epicyclic approximation by
$$\frac{\sigma_\phi}{\sigma_r} = \frac{\kappa}{2\Omega}.$$

Since the MUSE data have further confirmed that the velocity dispersion profile along the major axis of the host galaxy is flat with radius (see Fig.~\ref{kin_conf}), we have adopted
here the constant value  $\sigma_r$= 65 km/s.  For the $z$ dispersion, we simply adopt
$$ \sigma_z^2 = 2\pi Gh \Sigma(r),$$ 
\noindent where $\Sigma$ is the disk surface density.

Since neither the host disk nor the polar disk are exactly edge-on, we
projected them between 5$^\circ$ and 15$^\circ$ from edge-on, and the
polar disk is not quite perpendicular to the host but inclined by
20$^\circ$ from the pole.

\begin{figure*}
\centering
\includegraphics[width=18cm]{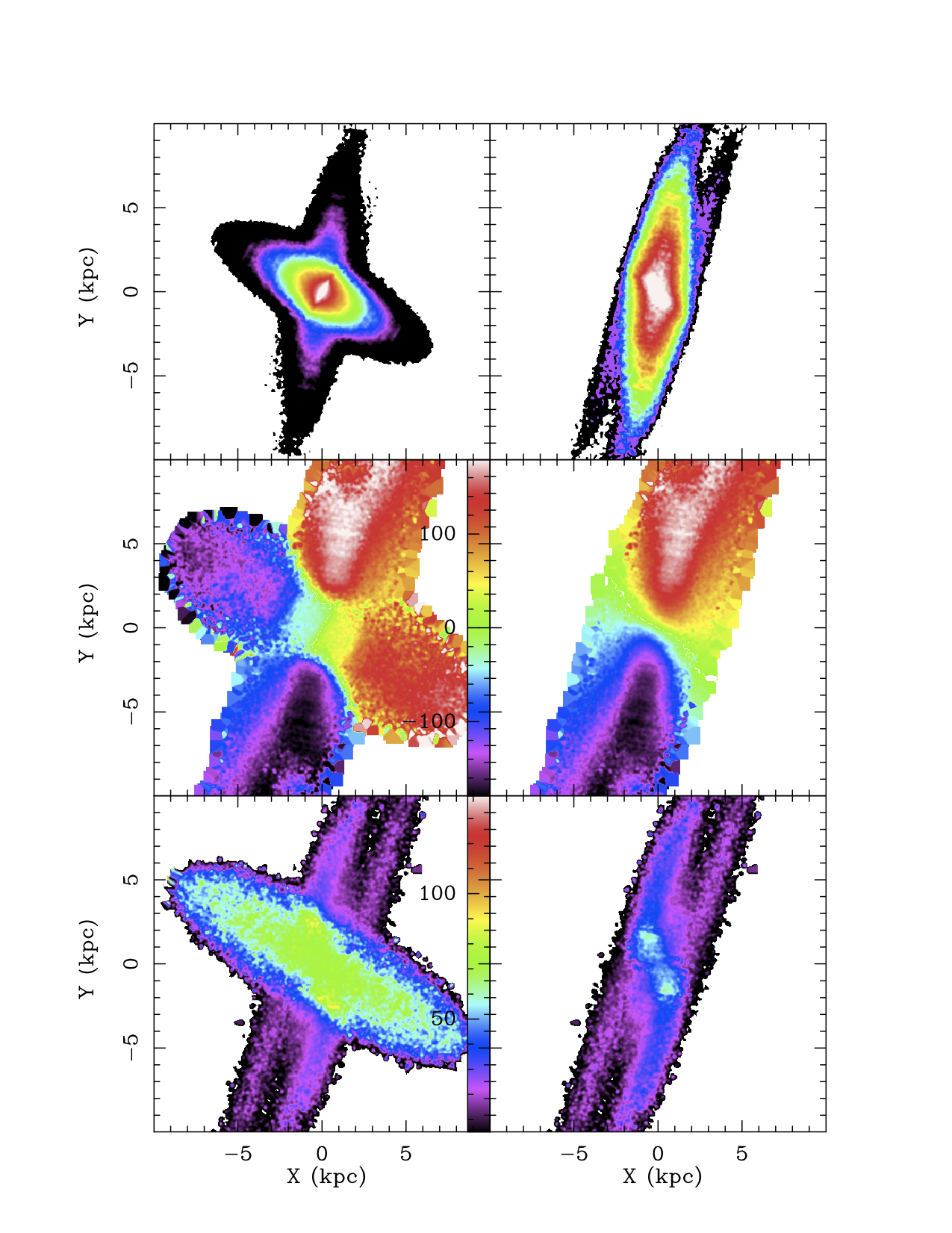} 
\caption{Projection of the model of the stellar (left panels) and 
    ionized gas kinematics (right) for NGC~4650A. The surface density
  distribution of stars and gas are shown in the upper panels, the
  projected velocity  in the middle panels, and the velocity
  dispersion in the bottom panels. The colour scales are indicated in
  km/s. North is up, east is on the left.}
\label{2Dmodel}
\end{figure*}

 The model was convolved with the seeing and the final result is
  shown in Fig.~\ref{2Dmodel} and in the bottom left panel in Fig.~\ref{zoom}.  
  
  The main result of this analysis is that the above
  model can account for the overall 2D field of the velocity and
  velocity dispersion derived by the MUSE data (see Fig.~\ref{2Dstar}
  and Fig.~\ref{2Dgas}). In particular, the line-of-sight velocities
  of stars and ionized gas obtained by the model show the same pattern
  and values as those observed in both HG and polar disk.
  In the region of the central host disk, the velocity dispersion of
  the stars (see bottom left panel in Fig.~\ref{2Dmodel}) is nearly
  constant at all radii, with comparable values
  ($40 \le \sigma \le 70$~km/s) to those observed in the MUSE data
  (see right panel in  Fig.~\ref{2Dstar}).

  We noticed some differences between the data and the model. 
In the region of the polar disk, the velocity dispersion of
  the stars and the gas derived from the model does not completely match
  that observed in the data. The model shows a lower stellar velocity
  dispersion ($\sim 30$~km/s, see bottom left panel in  Fig.~\ref{2Dmodel}) than that observed in the MUSE data
  ($\sim 100 - 150$~km/s, see right panel in  Fig.~\ref{2Dstar}). We noticed an increase in the velocity dispersion 
  of the model only in a restricted region at about 3~kpc (14~arcsec)
  on the NE and SW sides along the polar disk, where
  $\sigma \sim 90$~km/s, followed by a steep decrease. 

  The velocity dispersion of the ionized gas derived from the model
  shows a gradient (see bottom right panel in Fig.~\ref{2Dmodel}). It
  is higher in the centre ($\sim 50$~km/s for $R \le 2$~kpc
  $\sim 10$~arcsec) and then it decreases at larger radii, where
  $\sigma \sim 20 - 40$~km/s for $2 \le R \le 10$~kpc
  ($\sim 10-50$~arcsec). This gradient is not observed in the MUSE
  data, where the velocity dispersion of the gas along the polar disk
  is almost constant at all radii and $\sigma \sim 40$~km/s (see
  Fig.~\ref{2Dgas}, right panel). 

  Such differences in the velocity dispersion of stars and gas 
  could arise because the real structure of the polar disk
  is more complicated than that assumed in the simple geometrical
  model described above. In particular, one of the dominant features is
  the presence of the spiral arms, which are not included in the
  model. Moreover, the data show that the polar disk is characterized
  by several star forming regions that dominate the emissions along
  the spiral arms (see Fig.~\ref{mosaic}). Hence, the outflows from
  these regions may influence the measured velocity dispersion of the
  gas. This can explain why in the MUSE data $\sigma$ has a constant value at
  $\sim 40$~km/s at all radii, while in the model it decreases with radius to
  $\sim 20$~km/s.

  Finally, in the nuclear regions ($R\le 2$~arcsec $\sim 0.4$~kpc),
  the model does not show the apparent counter-rotation or the
  high velocity dispersion observed in the gas rotation curve and
  velocity dispersion profiles (see right panel in  Fig.~\ref{kin_gas}).  Thus, as suggested in Sect.~\ref{gas_center},
  this could be due to the presence of a real decoupled component or
  to the chaotic motions of the gas in the nuclear regions, also
  generated by an outflow, and the two features are not included in the
  model. Alternatively, the observed feature in the nuclear regions
  could be due to non-axisymmetric distortions generated by the two
  disks in the two planes. In fact, the model shows that none of
  the two disks have circular motions, the potential is never
  axisymmetric  in the host disk nor in the polar disk. This
  non-axisimmetry, combined with a position angle that is  not completely
  perpendicular, leads to non-circular motions that are seen tilted in
  projection (like an undetected bar misaligned from the major/minor
  axis).

  In the central regions of the HG, the model does not show the increasing
  velocity along the HG minor axis (see bottom left
  panel in Fig.~\ref{zoom}, as observed in the MUSE data (see top left
  panel in Fig.~\ref{zoom}). Along the NW-SE directions, the systemic
  velocity contour  in the stellar velocity field of the
  model is parallel to the HG minor axis for $R \le 2$~kpc
  ($\sim 10$~arcsec, see middle left panel in
  Fig.~\ref{2Dmodel}). Since the model does not include any rotation
  along this direction, this could be a real feature in the structure
  of the HG disk, and it  does not result from the geometry of the two disks and its projection on the sky.
 This confirms  the previous results by \citet{Coccato2014}. A definitive conclusion on this point
  will come from the 2D spectral decomposition of the
  MUSE stellar velocity field, which is the subject of a forthcoming  paper.

\begin{table}
      \caption[]{Masses and scales for the model components}
\label{tab:model}
\begin{center}
\begin{tabular}{lccccc}
\hline
\scriptsize{Component}& M                &  h    & r & $r_1$ & $r_2$ \\
                      &  10$^9$ M$_\odot$ & [kpc] & [kpc] & [kpc] & [kpc]  \\
               (1)    &  (2) & (3) & (4) & (5) & (6)\\
\hline
\smallskip
  HG bulge      &  $M_B$ &    &   $r_B$  &  & \\
                &  0.2 &      &   0.17 &   & \\
  HG disk       &  $M_d$ &   $h_d$ &   $r_d$ & & \\
                &  10.3 &   0.5 &   0.948   & & \\
  Polar disk       &  $M_{PD}$ &   $h_{PD}$ & &  &  \\
                   &  15.     &   0.5     & & 5.95  & 6.8   \\
  HI disk       &  $M_{HI}$ &   $h_{HI}$ &  &  &  \\
                &  7.2     &   0.5     &  & 3.4  & 15.3  \\
  DM halo       &  $M_{DH}$ &   $z$ &  $r_{h}$  &  &  \\
                &  15 &   1.2 &   6.0  &  &  \\
\hline
\end{tabular}
\tablefoot{Col.~1: Component included in the mass model. Col.~2: Total mass of each component. Col.~3: Characteristic scale-height of each component. Col.~4: Characteristic scale-length of each component. Col.~5 and Col.~6: $r_1$ and $r_2$ are the internal and external radii of the stellar and gaseous polar disks.}
\end{center}
\end{table}

\section{Summary and conclusions}


In this paper we presented new MUSE data obtained for the polar disk
galaxy NGC~4650A. The 2D velocity distribution of stars and
  ionized gas is mapped all the way to the galaxy centre where the
  two components co-exist, overcoming the limited spatial coverage of
  the long-slit data available for this object. They confirm that NGC~4650A is made of two perpendicular disks that rotate on  orthogonal planes, and that drive the kinematics right into the very centre of this galaxy.  
In the following we summarize the main results obtained by the
2D map of the line-of-sight velocity and velocity
dispersion of stars and  ionized gas, and we address the main
implications on the structure and formation of NGC~4650A.

\begin{itemize}

\item The final MUSE mosaic of NGC~4650A covers an area of
  $1.5 \times 2.5$~arcmin (see Fig.~\ref{mosaic}).  The MUSE image of
  the emission lines (see right panel in Fig.~\ref{mosaic}) confirms
  that  all the gas is  associated with the polar disk. Strong emissions are observed both in
  the outer arms and close to the centre of the galaxy.

\item  Overall, the velocity field of stars inside the HG has the
    typical pattern of a rotating disk, with receding velocities on
    the SW side and approaching velocities on the NE side. 
    In addition, we also observed that {\it i)} the velocity field is  not symmetric with respect to the centre, since the rotation curve   is flatter on the SW side, while it is still rising on the NE side
    (see Fig.~\ref{zoom}); and {\it ii)} there is a velocity gradient along the photometric minor axis of the HG.

\item  The velocity dispersion remains  almost constant, at
    $\sigma \sim 50 - 60$~km/s, at all distances from the galaxy
    centre and P.A.s inside the HG, while it increases along the polar
    disk, up to $\sigma \sim 100$~km/s in the outer regions.

\item Inside the HG regions, the photometric and kinematic
  P.A.  differ by about 5 degrees (see Fig.~\ref{PA}), with the
  photometric P.A., $P.A._{pho}=67 \pm 2$~degrees, is smaller than the
  kinematic P.A. 

\item  The 2D line-of-sight velocity of the ionized
    gas, which is distributed along the polar disk of NGC~4650A,
    (Fig.~\ref{2Dgas}, left panel) is mapped up to 75~arcsec
    ($\sim 16$~kpc) from the centre. It shows a large amount of
    differential rotation (of about 100 -- 120~km/s) from the centre
    up to the outer regions. On the NW and SE sides of the polar disk,
    the velocity of the approaching and receding arms is measured,
    including the regions where they overlap with the underlying disk. 

\item The gas velocity dispersion remains almost constant
  ($\sigma_{gas} \sim 30 - 40$~km/s) at all radii (see right panel in  Fig.~\ref{2Dgas}), except inside 2~arcsec from the galaxy centre,
  where we measured a steep increment of $\sigma$ that reaches a value
  of about 110~km/s. In the same range of radii, we also observe a
  hint of a decoupled component in the velocity curve (see right
  panel in Fig.~\ref{kin_gas}).

\end{itemize}

 In order to reproduce the observed structure and the new 2D
  kinematics for NGC~4650A given by the MUSE data, we constructed a
  multicomponent mass model made by the combined projection of two
  disks, plus a dark matter halo (see Sect.~\ref{model}). It is based
  on the dynamical model developed by \citet{Combes1996}, where the dark halo is flattened and  its major axis is aligned with the polar disk. The structural parameters for the light distribution take 
   the extended and deeper photometry published by
  \citet{Iod02} into account. It is a ``minimum'' model that does not include the
  spiral arms in the polar disk, the rotation along the minor axis of the HG and
  the nuclear decoupled component in the centre. 

  By comparing the observations with the 2D kinematics derived by the
  model, we find that the complex velocity field revealed by the MUSE
  data for NGC~4650A is overall well reproduced by this simple model,
  also in the central regions of the galaxy, where the two components
  coexist (see Fig.~\ref{2Dmodel}).

This result is a strong constraint on the dynamics and formation history of the polar disk, in particular on the nature of the dark matter. 
In the framework of disk galaxy formation, an extended and massive polar disk like that observed in NGC~4650A can form through the accretion of cold gas along a filament. In this scenario, the central HG is disk that formed first. The decoupling of the angular momentum at a certain epoch of the formation history moved the accretion of gas in the orthogonal direction, forming the polar disk \citep{Snaith12,Combes2014}. Alternatively, a polar disk could also form from the tidal accretion of material from a gas-rich companion galaxy. Both mechanisms were successfully tested for  NGC~4650A  \citep{Spav10}.

The analysis performed in this work shows that a mass model made by the two perpendicular disks  reproduce well the observed structure and kinematics of NGC~4650A, which result from one of the accretion mechanism described above. Since the mass model includes a flattened dark halo, having its major axis aligned with the polar disk, this suggests that if the polar disk is  formed through the accretion of material from outside,  dissipative dark matter might to be essential.

  Some differences are noted between the observed and modelled
  kinematics. They are the rotation velocity along the minor axis
  of the HG, the high stellar velocity dispersion in the
  polar disk, and a possible decoupled component in the galaxy
  centre, which are detected in the data but not in the model. 

  They do not affect the main conclusion discussed above, since they
  are limited to regions that are not dynamically important for the
  whole model. In fact, inside the HG, the minor axis rotation is for
  $R\le 5$~arcsec ($\sim 1$~kpc), which are only  about  $15\%$
  of the measured kinematics along the major axis, and which are fully reproduced by the model.

  In the polar disk, we measured an increasing velocity dispersion of
  stars, for $5 \le R \le 10$~arcsec ($\sim 1 - 2$~kpc), out to about
  110~km/s (see right panel in Fig.~\ref{2Dstar}), and a difference in
  the velocity dispersion of the gas of about 20~km/s at 16~kpc (see
  right panel in Fig.~\ref{2Dgas} and right bottom panel in  Fig.~\ref{model}). Along the polar disk, the dynamics is strongly
  constrained by the high velocities ($\sim 120$~km/s) of the H{\small I} gas at 40~kpc from the centre.

  As pointed out in Sect.~\ref{model}, such differences in both stellar
  and gaseous velocity dispersion could be a consequence of the
  perturbed structure of the polar disk, which are not taken into
  account by the simple model made in this work. The differences in
  the measured velocity dispersion of the gas with respect to that
  derived from the model could be due to the outflows from the several
  star forming regions that dominate the emissions along the spiral
  arms. On the other hand, the higher velocity dispersion of the stars
  in the outer regions of the polar disk and the counter-rotation in
  the centre of the polar disk, if the latter is a real decoupled
  component, could be the result of the angular momentum evolution of
  the accreted material during the formation process.


\begin{acknowledgements}
  This work is based on observations taken at the ESO La Silla Paranal
  Observatory within the MUSE Commissioning.   The authors wish to
    thank the anonymous referee for his/her comments and suggestions
    that allowed us to greatly improve  the paper. E.I. wishes to thank the
  European Southern Observatory for the financial support, the
  hospitality and the access to the computer facilities during the
  several visits in 2014 and 2015 that allowed us to perform the 
  fundamental steps of the work presented in this
  paper. F.C. acknowledges the European Research Council for
 the Advanced Grant Program Num 267399-Momentum. P.M.W. received funding through BMBF Verbundforschung (project
  MUSE-AO, grant 05A14BAC and 05A14MGA). R.B. acknowledges support from the ERC advanced grant
  339659-MUSICOS.
\end{acknowledgements}



\bibliography{n4650a}

\end{document}